\documentclass[a4paper,11pt]{article}
\pdfoutput=1 % if your are submitting a pdflatex (i.e. if you have
\usepackage{jheppub} % for details on the use of the package, please
\usepackage[T1]{fontenc} % if needed

\title{Quasinormal Modes of Charged Black Holes Localized in the Randall-Sundrum Brane World}

\author[a]{N. Abbasvandi,}
\author[b,c,1]{M. J. Soleimani,\note{Corresponding author.}}

\author[b,c]{W.A.T. Wan Abdullah}
\author[a]{Shahidan Radiman}

\affiliation[b]{School of applied Physics, Faculty of Science $\&$ Technology, Universiti Kebangsaan Malaysia, 43600 Bangi, Malaysia}
\affiliation[a]{Department of Physics, Faculty of Science, University of Malaya, 50603 KL, Malaysia}
\affiliation[c]{National Center for Particle Physics, IPPP, University of Malaya, 50603 KL, Malaysia}

\emailAdd{msoleima@cern.ch}
\emailAdd{Niloofar@siswa.ukm.edu.my}
\emailAdd{wat@um.edu.my}
\emailAdd{shahidan@um.edu.my}

\abstract{We study the quasinormal modes of the massless scalar field of charged black holes embedded in the Randal-Sundrum brane world using the third order WKB approximation. We consider the effects of the electromagnetic and tidal charges on quasinormal frequencies spectrum for charged black holes as well as the effect of the thickness of the bulk.}

\begin{document}
	\maketitle
	\flushbottom
	
	\section{Introduction}
	The quasinormal modes (QNMs) of black holes \cite{vishveshwara,chandrasekhar,kokkotas,nollert,konoplya,berti,konoplya1}, present a discrete set of complex frequencies corresponding to the solutions of the perturbation equations. The QNMs depend only on black hole parameters but not depend on the initial perturbation. In general, the imaginary part of the complex frequencies are associated with the decay timescale of the perturbation while the real part represents the actual frequency of the oscillation. During last decades, a lot of authors had focused on the QNMs of black holes in asymptotically flat spacetime \cite{berti2003,berti20031,berti2004}, asymptotically (anti-)de Sitter (henceforth AdS or dS) spacetimes \cite{mann97,mann98,kalyanarama,danielsson,danielsson99,brady,abdalla}, scalar perturbation \cite{wang,chakrabarti}, gravitational perturbation \cite{giammatteo}, Dirac perturbation \cite{cho,jing04,jing05}, and electromagnetic perturbation \cite{lopez}, in different backgrounds.
	The investigation of QNMs has been extended to higher dimensional spacetime (e.g. \cite{chakrabarti,lopez}). Indeed, the QNMs contain information on the parameters of the underlying black hole and hence have significance in identifying black holes in four and higher dimensions. 
	Recently, the possiblity of production of TeV-scale black holes \cite{argyres,bank,emparan,giddings} at particle colliders such as the Large Hadron Collider (LHC) have attracted much attention to the models of low scale gravity \cite{arkani,antoniadis,randall1,randall2}. Black holes in higher dimensional spacetime have attracted a lot of interest within different theories such as supersymmetric string theory and different extra dimensional scenarios namely Arkani-Hamed, Dimopoulos and Dvali (ADD) \cite{arkani} and Randall-Sundrum \cite{randall1,randall2} (RS) brane world models (for a review see \cite{kanti}). In this way, based upon the fact that our universe may be embedded in higher dimensional spacetime, much interest has been paid to the RS model to resolve the gauge hierarchy problem. In the context of the RS brane world scenario, there has been an attempt to construct numerical black holes, being successful only in working out small localized black holes \cite{kudoh} and some remarks were made in terms of the AdS/CFT correspondence as to why finding black hole solution localized on the RS brane is so difficult that large black holes will not be static \cite{tanaka,fabbri}. Recently, in a pioneering work the role of black hole quasinormal modes in gravitational
	experiments devised to determine the existence of the brane, in a lower dimensional setting \cite{chung}.
	\\In this paper, we discuss the quasinormal modes (QNMs) of charged brane world black holes in the Randall Sundrum model considering both the electromagnetic and tidal charges effects as well as the thickness of the bulk. There are several techniques to compute the QNMs, such as the potential fit \cite{ferrari}, Leaver's continued fraction method \cite{leaver}, Wentzel-Kramers-Brillouin (WKB) \cite{schutz,iyer}, etc. Here, we study the QNMs of charged black holes embedded on the RS brane world considering quantum gravity effects and using the 3th order WKB approximation as the results do improve for higher orders of WKB \cite{konoplya}. We discuss in detail how the QNMs are influenced by the parameters of higher dimensional black hole on bulk spacetime.
	\\This paper is organized as follows: In section II we set up the basic notion and formalism of the charged black hole localized on the RS brane world and we obtain the effective potential, solving 5-dimensional Klein-Gordon wave equation. In section III we compute quasinormal modes of the charged black hole in brane world using 3th order WKB approximation. In the last section a summary of results, conclusions and discussion are given.    
	
	\section{Charged Brane World Black Hole}
	In order to obtain the charged black hole solution localized on the Randall-Sundrum (RS)\cite{randall1,randall2} brane world, we start with the action of the RS model in (4+1) dimensions as follows
	\begin{equation} \label{1}
	{S_5} = \frac{1}{{16\pi G_N^{\left( 5 \right)}}}\int {{d^4}x\int {d\chi\sqrt { - {g_{\left( 5 \right)}}} \left[ {{R^{\left( 5 \right)}} + 12{\kappa ^2}} \right]}  - \int {{d^4}x\left[ {\sqrt { - {g_ + }} {\lambda _ + } + \sqrt { - {g_ - }} {\lambda _ - }} \right]} }
	\end{equation} 
	where $3{\kappa ^2}$ is a cosmological constant while $\lambda _ +$ and $\lambda _ -$ are tensions of the branes at $\chi_1 = 0$ and $\chi_2 = \pi {r_c}$ respectively which are two singular points on the orbifold ${{{S^1}} \mathord{\left/{\vphantom {{{S^1}} {{Z_2}}}} \right.\kern-\nulldelimiterspace} {{Z_2}}}$ located at ${\chi _1}$ and ${\chi _2}$. Here, two 3-branes are placed at these points and we assume that orbifold possesses a periodicity in the extra coordinate $\chi$, identifying $-\chi$ with $\chi$. In this case, one can define the metric on each brane as
	\begin{equation} \label{2}
	\begin{array}{l}
	g_{\mu \nu }^{\left(  +  \right)} \equiv g_{\mu \nu }^{\left( 5 \right)}\left( {{x^\mu },\chi = 0} \right)\\
	g_{\mu \nu }^{\left(  -  \right)} \equiv g_{\mu \nu }^{\left( 5 \right)}\left( {{x^\mu },\chi = \pi r} \right).
	\end{array}
	\end{equation} 
	We assume that the induced metric on the brane take the general form 
	\begin{equation}
	d{s^2} =  - f\left( r \right)d{t^2} + \frac{{d{r^2}}}{{f\left( r \right)}} + {r^2}d\Omega _2^2
	\end{equation}
	where $d\Omega _2^2 = d{\theta ^2} + {\sin ^2}\theta d{\varphi ^2}$. In this manner, one can assume the bulk metric take the following form 
	\begin{equation} \label{3}
	d{s^2} = {\gamma ^2}\left( {\chi ,r} \right)d{\chi ^2} - {e^{2\varsigma \left( {\chi ,r} \right)}}f\left( r \right)d{t^2} + {e^{2\upsilon \left( {\chi ,r} \right)}}{f^{ - 1}}\left( r \right)d{r^2} + {e^{2\tau \left( {\chi ,r} \right)}}{r^2}d\Omega _{\left( 2 \right)}^2
	\end{equation} 
	where $\gamma\left( \chi \right)$ is the lapse function which describes embedding geometry of the hyper surface spanned by ${\left( {t,r,\theta ,\varphi } \right)}$ during the evolution in the bulk spacetime and ${d\Omega _{\left( 2 \right)}^2}$ is a metric of unit $2$-sphere. Among possible solution satisfying the equation of motion \cite{chamblin}, we consider charged black hole solution in the RS model in which our universe is viewed as a domain wall in asymptotically AdS space. In this case, using the Hamiltonian constraint and solving Hamiltonian equation along with metric ansatz (2.3), one can obtain \cite{chamblin}
	\begin{equation} \label{4}
	f(r) = 1 - \frac{{2{G_4}M}}{r} + \frac{{{Q^2} + \zeta }}{{{r^2}}} + \frac{{l_p^2{Q^4}}}{{20{r^6}}}
	\end{equation} 
	Here, $M$ is the mass of black hole. $\zeta$ and $Q$ are tidal and electromagnetic charges respectively and we use $r_+$ to denote the position of the outermost horizon which ${f\left( {{r_ + }} \right) = 0}$. We draw attention, the only apparent horizon that appears during the $\chi$ evolution is located at $r = r_+$. It was shown in the Appendix of \cite{shinkai} that at $r = r_+$, $\varsigma$ and $\upsilon$ evolve synchronously, that is , $\varsigma \left( {T,{r_ + }} \right) = \upsilon \left( {T,{r_ + }} \right)$ and analytically continuing back to original spacetime yields $\varsigma \left( {\chi ,{r_ + }} \right) = \upsilon \left( {\chi ,{r_ + }} \right)$. Therefore, here for simplicity, we set $\varsigma \left( {\chi ,r} \right) = \upsilon \left( {\chi ,r} \right) = \tau \left( {\chi ,r} \right) =  - \kappa \chi$. Then, we obtain
	\begin{equation} \label{5}
	d{s^2} = {e^{ - 2\kappa \chi }}\left[ { - f\left( r \right)d{t^2} + {f^{ - 1}}\left( r \right)d{r^2} + {r^2}d\Omega _{\left( 2 \right)}^2 ]+ d{\chi ^2}} \right.
	\end{equation}
	Indeed, a Randall Sundrum type brane model is built up which has two branes at $\chi = 0$ and $\chi = \pi r_c$. The solution (2.6) describes a charged black hole placed on the hypersurface at the fixed extra coordinates which in this brane background, the scalar field with mass $m$, satisfies the Klein-Gordon wave equation as
	\begin{equation} \label{6}
	\left( {\nabla _{\left( 5 \right)}^2 - {m^2}} \right)\Phi  = 0
	\end{equation}
	which one can write it explicitly as follow
	\begin{equation} \label{7}
	\begin{array}{l}
	{e^{2\kappa \chi}}\left[ {\frac{{ - 1}}{f}{\partial _{,tt}}\Phi  + \frac{1}{{{r^2}}}{\partial _r}\left( {{r^2}f{\partial _r}\Phi } \right) + \frac{1}{{{r^2}\sin \theta }}{\partial _\theta }\left( {\sin \theta {\partial _\theta }\Phi } \right) + \frac{1}{{{r^2}{{\sin }^2}\theta }}{\partial _{,\varphi \varphi }}\Phi } \right]\\
	\begin{array}{*{20}{c}}
	{}&{ + {e^{4\kappa \chi}}{\partial _\chi}}
	\end{array}\left( {{e^{ - 4\kappa \chi}}{\partial _\chi}\Phi } \right) - {m^2}\Phi  = 0
	\end{array}
	\end{equation}
	with $\Phi \left( {t,r,\theta ,\varphi ,\chi} \right) = \psi \left( {t,r,\theta ,\varphi } \right)\xi \left( \chi \right)$.
	Here, the separation of variables is easily possible if we set
	\begin{equation} \label{8}
	{\partial _{,\chi \chi }}\xi  + 3{\kappa ^3}{\partial _\chi } - {m^2}\xi  + \lambda^2 \xi  = 0
	\end{equation}
	which is discussed in \cite{liu}. In this case, the modes along the extra dimension are quantized by stable standing waves, and then the eigenvalue is naturally discretized as 
	\begin{equation} \label{9}
	{\xi _n}\left( \chi  \right) = C{e^{\frac{{ - 3\kappa }}{2}\chi }}\cos \left( {n\pi \frac{\chi _1 }{{{\chi}}}} \right)
	\end{equation}
	where the quantum parameter $\lambda$ is
	\begin{equation} \label{10}
	{\lambda^2 _n} = \frac{{{n^2}{\pi ^2}}}{{\chi^2}} + \frac{9}{4}{\kappa ^2}
	\end{equation}
	where $n = 1,2,3,...$ and $\chi$ is the thickness of the bulk. In this way, the reduced form of the equation (2.7) reads
	\begin{equation} \label{11}
	\frac{{ - 1}}{f}{\partial _{,tt}}\psi  + \frac{1}{{{r^2}}}{\partial _r}\left( {{r^2}f{\partial _r}\psi } \right) + \frac{1}{{{r^2}\sin \theta }}{\partial _\theta }\left( {\sin \theta {\partial _\theta }\psi } \right) + \frac{1}{{{r^2}{{\sin }^2}\theta }}{\partial _{,\varphi \varphi }}\psi-{\lambda _n}\psi = 0.
	\end{equation} 
	We substitute four dimensional wave function resolving the field into spherical harmonics as follows
	\begin{equation} \label{12}
	\psi \left( {t,r,\theta ,\varphi } \right) = \sum\limits_{lm} {\frac{1}{r}{e^{ - i\omega t}}\psi _m^l\left( r \right)Y_l^m\left( {\theta ,\varphi } \right)}
	\end{equation}
	which is the usual separation of variables in terms of a radial field and a spherical harmonic $Y_l^m\left( {\theta ,\varphi } \right)$. In the region the submanifold given by the patch ${T_ + } = \left\{ {\left( {t,r,\theta ,\varphi } \right),r > {r_ + }} \right\}$ which we will treat the dynamics of fields in the black hole exterior, one can define a $tortoise$ $coordinate$ ${r^*}\left( r \right)$ as follows
	\begin{equation} \label{13}
	d{r^*} = \frac{{dr}}{{f\left( r \right)}}
	\end{equation} 
	where ${f\left( {{r_ + }} \right) = 0}$ and the black hole horizon located at $r_+$. In this case, the radial function of (2.13) satisfy a Schr\"{o}dinger wave equation which for each multiple moment reads
	\begin{equation} \label{14}
	\left( {{\partial _{,{r^*}{r^*}}} + {\omega ^2} - V\left( {{r}} \right)} \right)\psi _m^l\left( {{r^*}} \right) = 0
	\end{equation}
	where $V\left( r \right)$ is the effective potential given by
	\begin{equation} \label{15}
	V\left( r \right) = \left( {1 - {\mkern 1mu} \frac{{2{G_4}M}}{r} + \frac{{{Q^2} + \zeta }}{{{r^2}}} + {\mkern 1mu} \frac{{{l_p}^2{Q^2}}}{{20{r^6}}}} \right) \times \left( {\frac{{L\left( {L + 1} \right)}}{{{r^2}}} + {\mkern 1mu} {\lambda^2 _n} + \frac{{2{G_4}M}}{{{r^3}}} - 2{\mkern 1mu} \frac{{{Q^2} + \zeta }}{{{r^4}}} - {\mkern 1mu} \frac{{3{l_p}^2{Q^2}}}{{10{r^8}}} - {m^2}} \right).
	\end{equation}
	Here, the spectrum of spherical harmonics $\Delta Y_l^m\left( {\theta ,\varphi } \right)$ is given by
	\begin{equation} \label{16}
	\Delta Y_l^m\left( {\theta ,\varphi } \right) =  - L\left( {L + 1} \right)Y_l^m\left( {\theta ,\varphi } \right).
	\end{equation}
	Here, the tortoise coordinate $r*$ is defined on the interval $\left( { - \infty , + \infty } \right)$ in such a way that the spatial infinity $r =  + \infty$ correspond to ${r^*} = \infty$, while the event horizon correspond to ${r^*} =  - \infty$. The effective potential (2.16) is positively defined in the region ${r^*} \in \left( { - \infty , + \infty } \right)$ and has the form of the potential barrier which approaches constant values at both spatial infinity and event horizon, i.e. outside the event horizon $r \in \left[ {{r_ + }, + \infty } \right)$. 
	\begin{figure}[t] \label{1}	\centering 
		\begin{center} 
			\centering
			\includegraphics[width=0.4\textwidth,trim=110 350 60 60,clip]{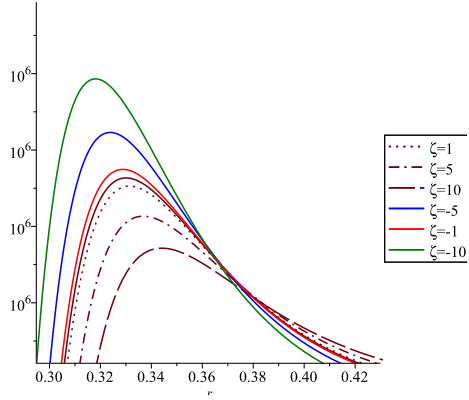}
			\caption{\label{fig:i}The effective potential $V\left( r \right)$ versus radial coordinate $r$ with $L = 0$ for different values of $\zeta$ in presence of the fixed $Q = 3$. Here, we adopt $M=100$, $n = \kappa = 1$, $\chi = 10$.}
		\end{center}
	\end{figure} 
	\begin{figure}[t] \label{2}	\centering 
		\begin{center} 
			\centering
			\includegraphics[width=0.4\textwidth,trim=110 350 60 60,clip]{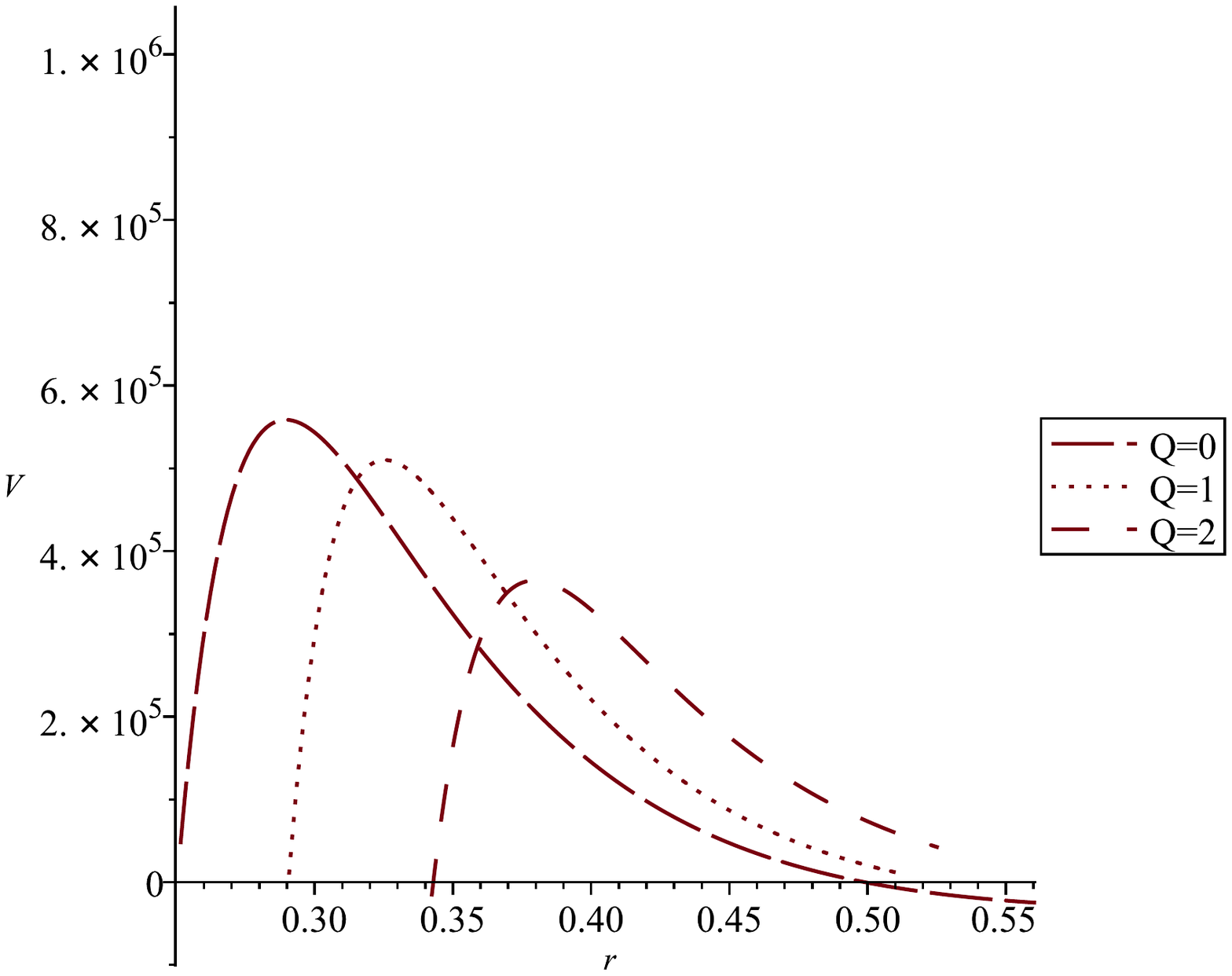}
			\caption{\label{fig:i}The effective potential $V\left( r \right)$ versus radial coordinate $r$ with $L = 0$ for different values of $Q$ in presence of the fixed $\zeta = 50$. Here, we adopt $M=100$, $n = \kappa = 1$, $\chi = 10$.}
		\end{center}
	\end{figure}  
	\begin{figure}[t] \label{3}	\centering 
		\begin{center} 
			\centering
			\includegraphics[width=0.5\textwidth,trim=100 70 10 50,clip]{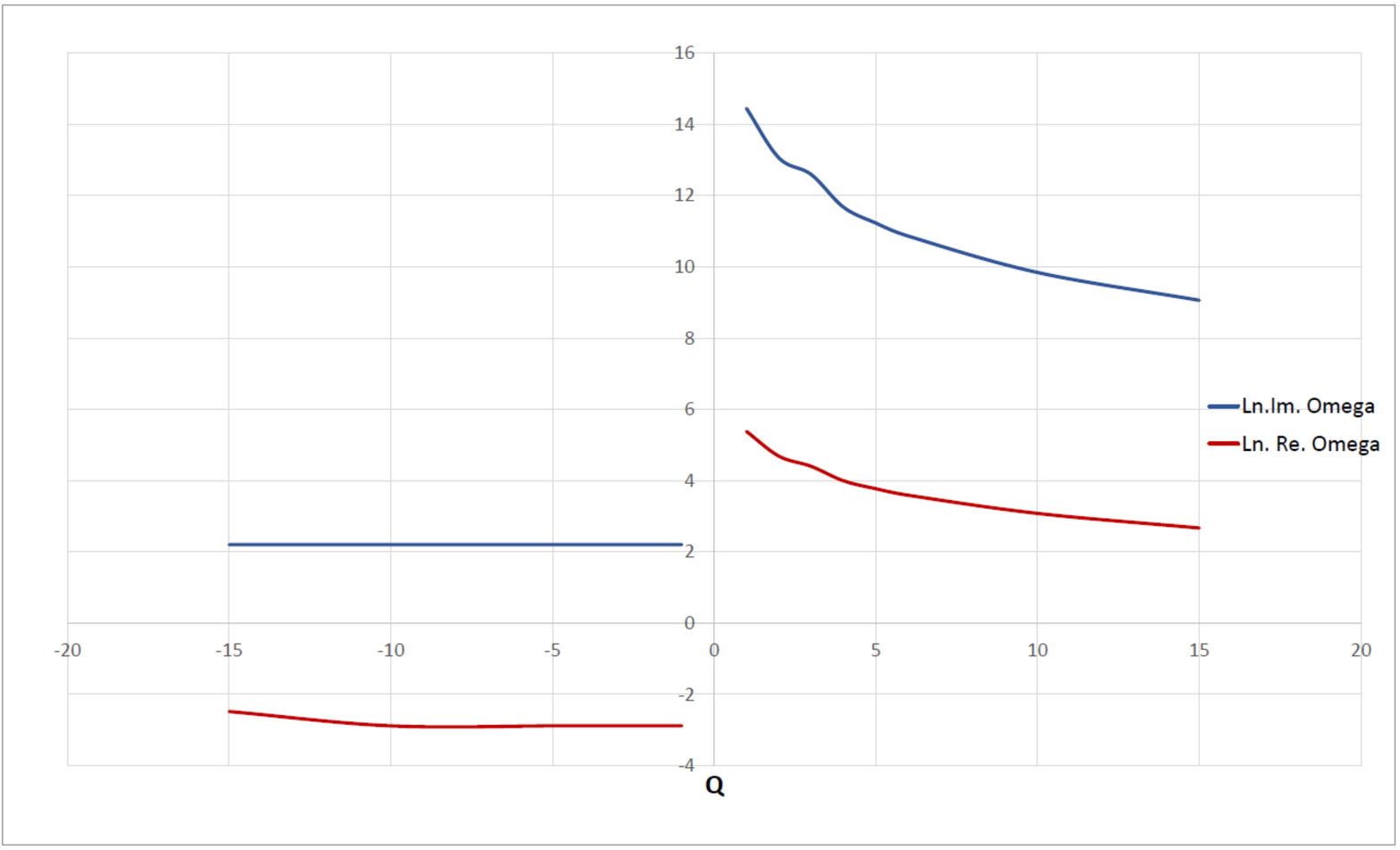}
			\caption{\label{fig:i}The logarithmic imaginary and logarithmic real part of the quasinormal frequencies in the absence of the electromagnetic charge, $Q = 0$, for various values of $\zeta$. We adopt $M=100$, $n = \kappa = 1$, $\chi = 10$.}
		\end{center}
	\end{figure}  
	\begin{figure}[t] \label{4}	\centering 
		\begin{center} 
			\centering
			\includegraphics[width=0.4\textwidth,trim=40 70 10 50,clip]{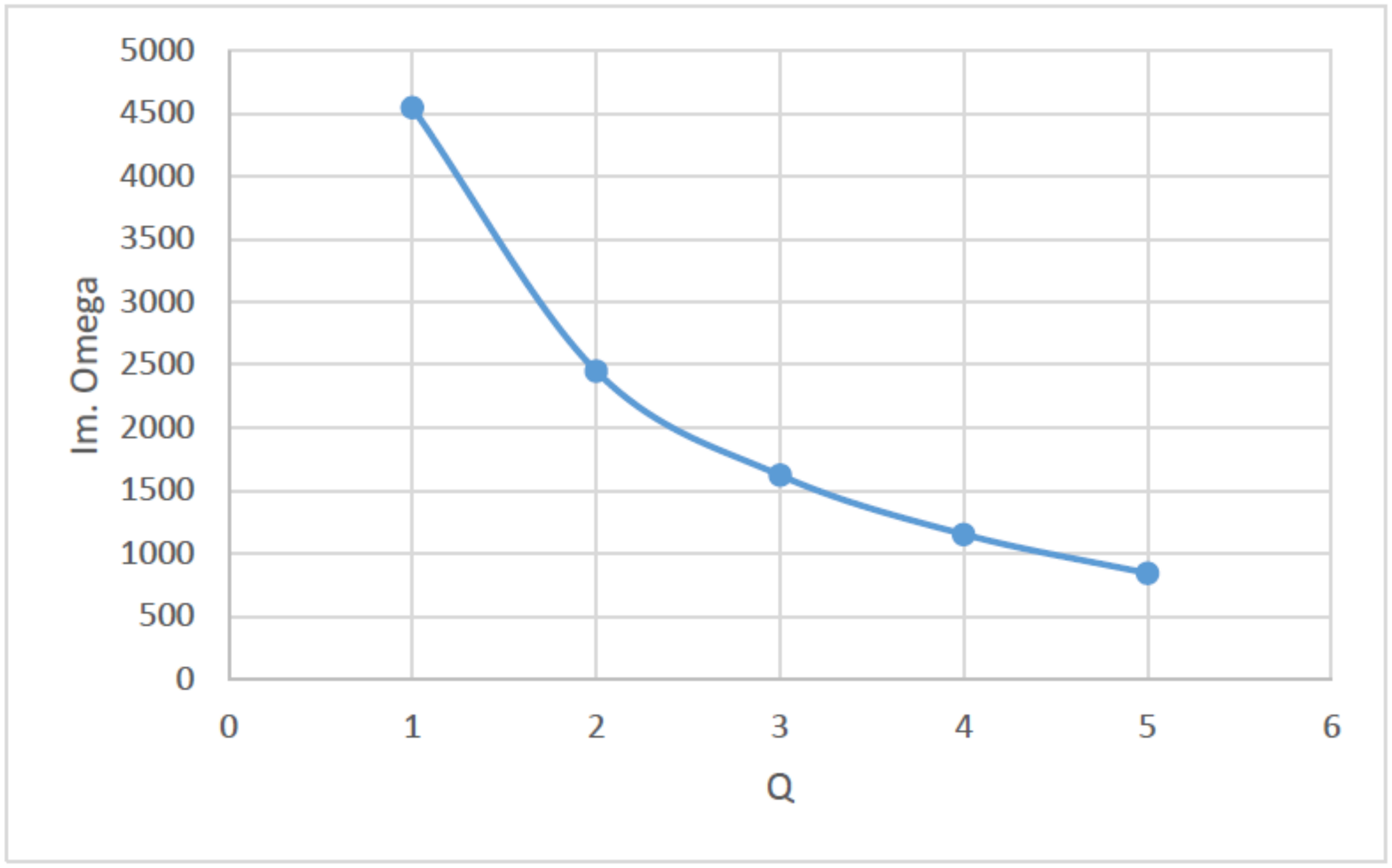}
			\includegraphics[width=0.4\textwidth,trim=40 70 10 50,clip]{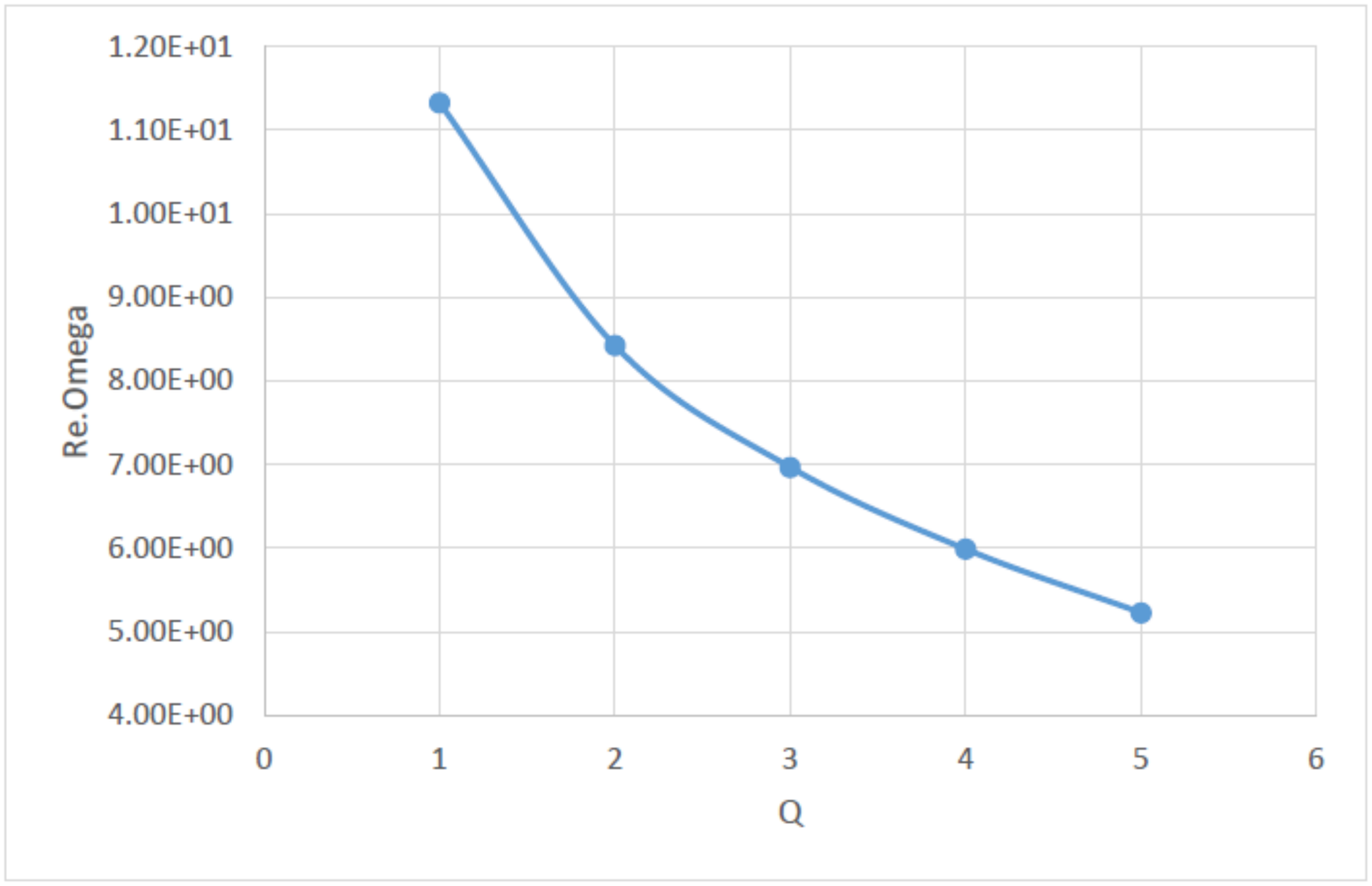}
			\caption{\label{fig:i}The imaginary part (left) and real part (right) of the $\omega$ with respect to $Q$ in the absence of the tidal charge, $\zeta = 0$ with adopted $M=100$, $n = \kappa = 1$, $\chi = 10$.}
		\end{center}
	\end{figure}  
	\begin{figure}[t] \label{5}	\centering 
		\begin{center} 
			\centering
			\includegraphics[width=0.4\textwidth,trim=30 70 10 50,clip]{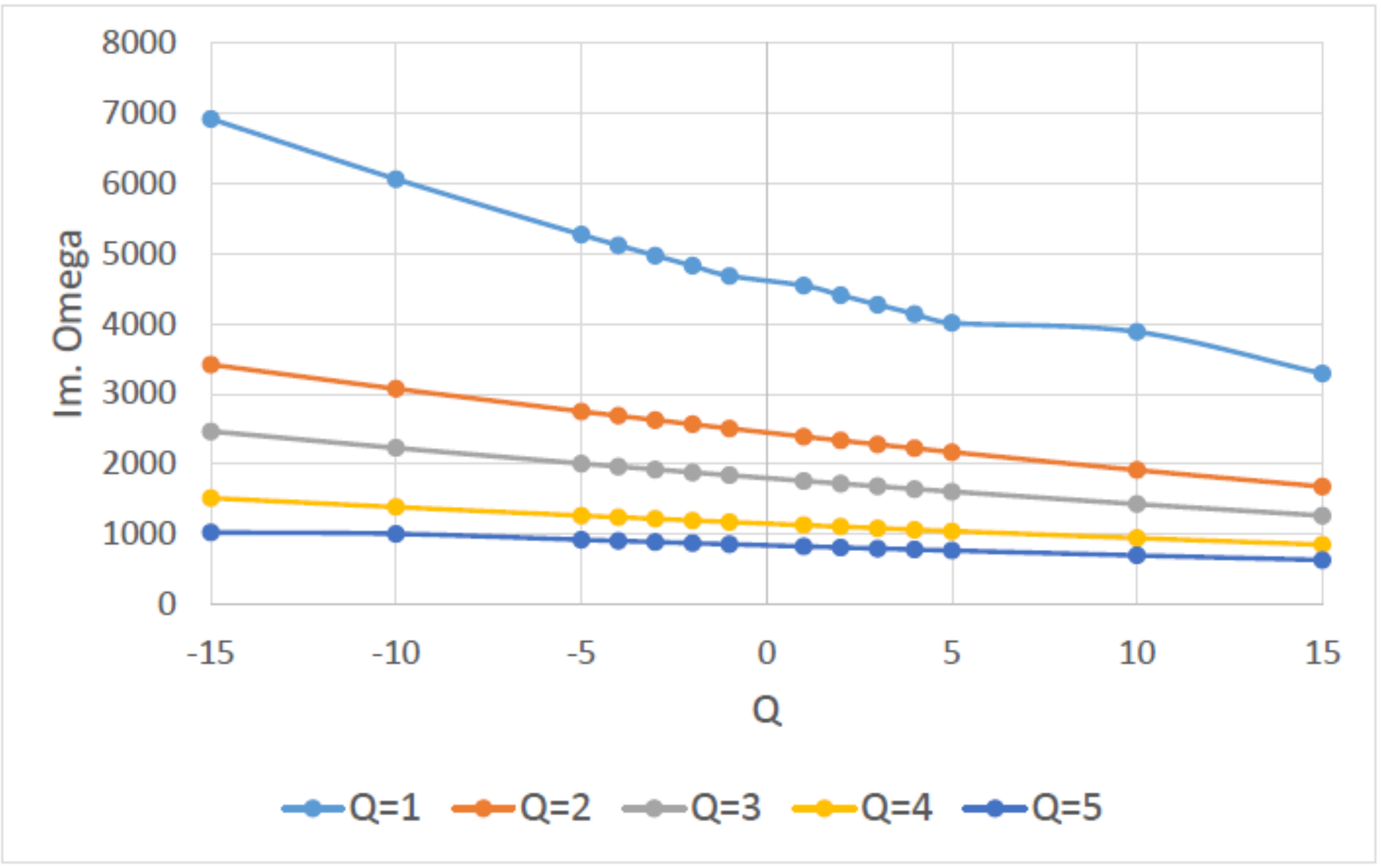}
			\includegraphics[width=0.4\textwidth,trim=30 50 10 50,clip]{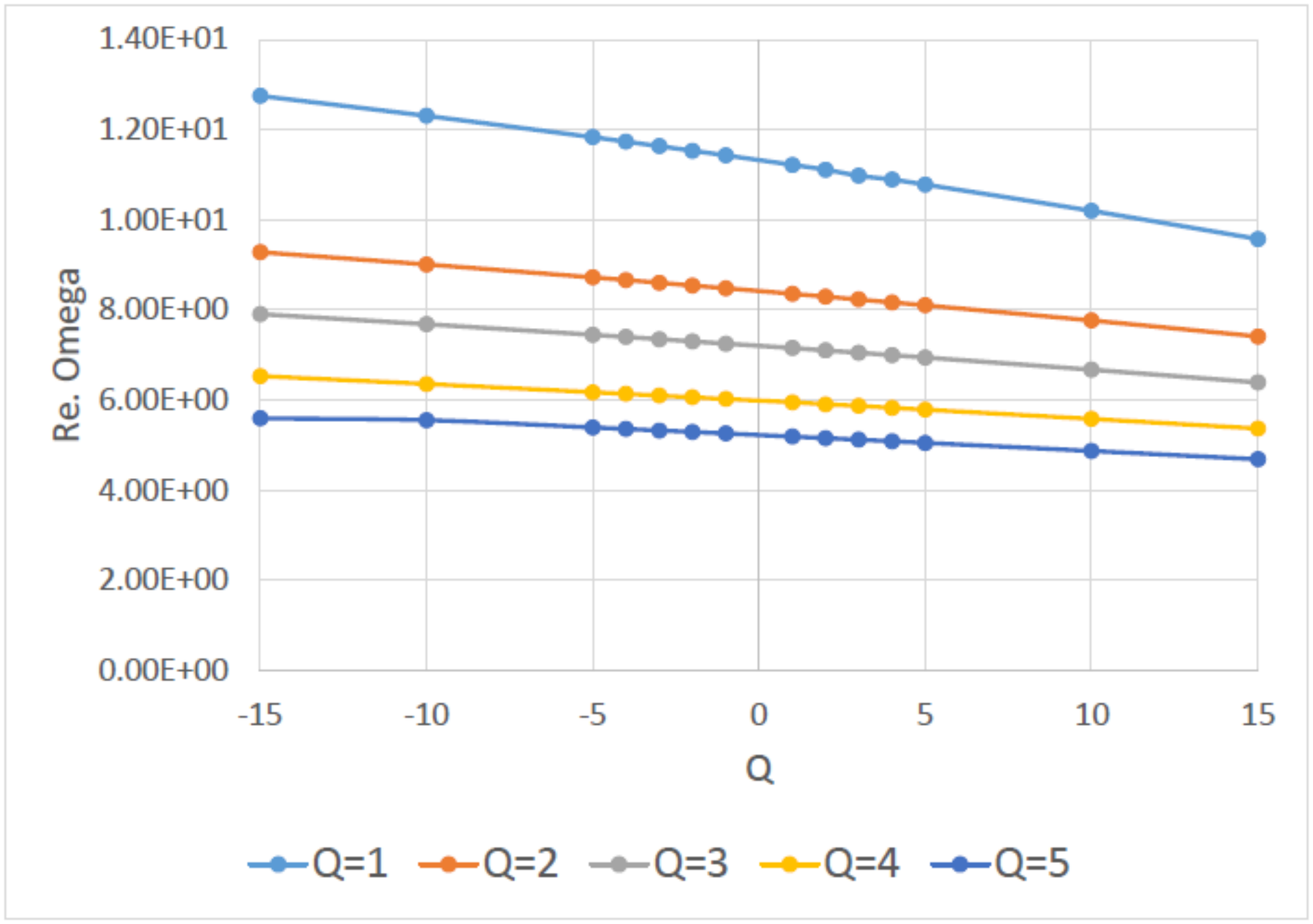}
			\caption{\label{fig:i}The imaginary part (left) and real part (right) of the $\omega$ with respect to the $\zeta$ for different electromagnetic charge, $Q$, with adopted $M=100$, $n = \kappa = 1$, $\chi = 10$.}
		\end{center}
	\end{figure}
	
	The five dimensional effective potential, $V\left( r \right)$, contains the quantum number $n$. In this case, $V\left( r \right)$ is higher and thicker than the four dimensional case when $\lambda_n = 0$. In this manner, the effective potential is related to the mass of the black hole, $M$, electromagnetic charge, $Q$, and tidal charge, $\zeta$. Here, we consider three cases: i) No electromagnetic charge, i.e., $Q = 0$ which the induced metric has a regular horizon if $\zeta  \le {M^2}$, ii) No tidal charge, i.e., $\zeta = 0$, which the induced metric on the domain wall is Reissner-Nordstr\"{o}m with a correction term \cite{chamblin}, and iii) Both charges non-zero, i.e., $\zeta \ne 0$, $Q \ne 0$. Figure 1 represent the typical variation of the effective potential $V\left( r \right)$ for different values of the tidal charges in presence of fixed value of the electromagnetic charge. It is obvious that increasing value of both tidal and electromagnetic charges causes decrease of the peak of the potential barrier (see Figure 1 and 2).  
	\begin{figure}[t] \label{6}	\centering 
		\begin{center} 
			\centering
			\includegraphics[width=0.5\textwidth,trim=30 70 10 50,clip]{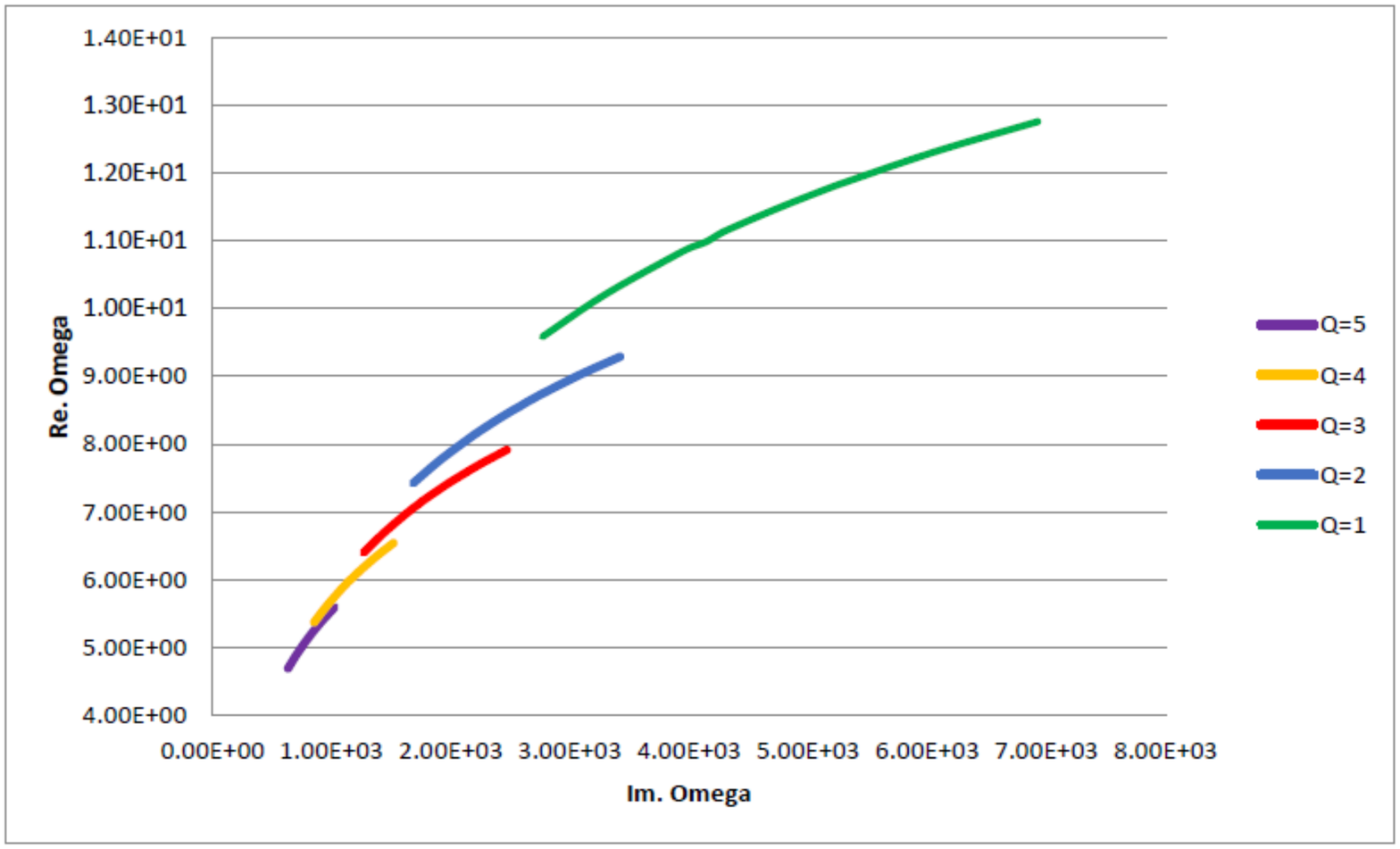}
			\caption{\label{fig:i}The real part of the $\omega$ (Omega) Vs. the imaginary part of the $\omega$ for various values of the electromagnetic charge, $Q$, We adopt $M=100$, $n = \kappa = 1$, $\chi = 10$.}
		\end{center}
	\end{figure}
	\begin{figure}[t] \label{7}	\centering 
		\begin{center} 
			\centering
			\includegraphics[width=0.4\textwidth,trim=30 70 10 50,clip]{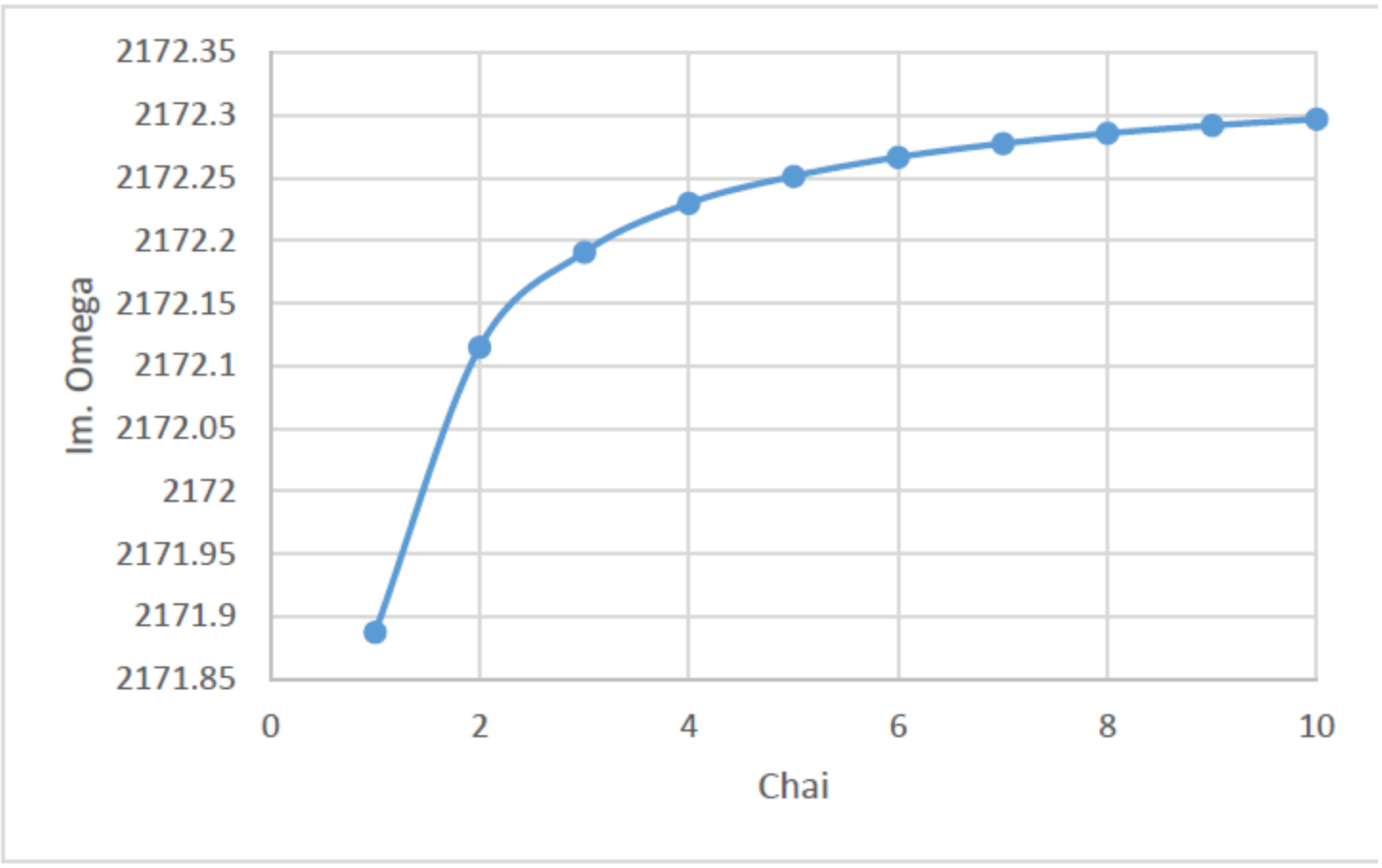}
			\includegraphics[width=0.4\textwidth,trim=30 50 10 50,clip]{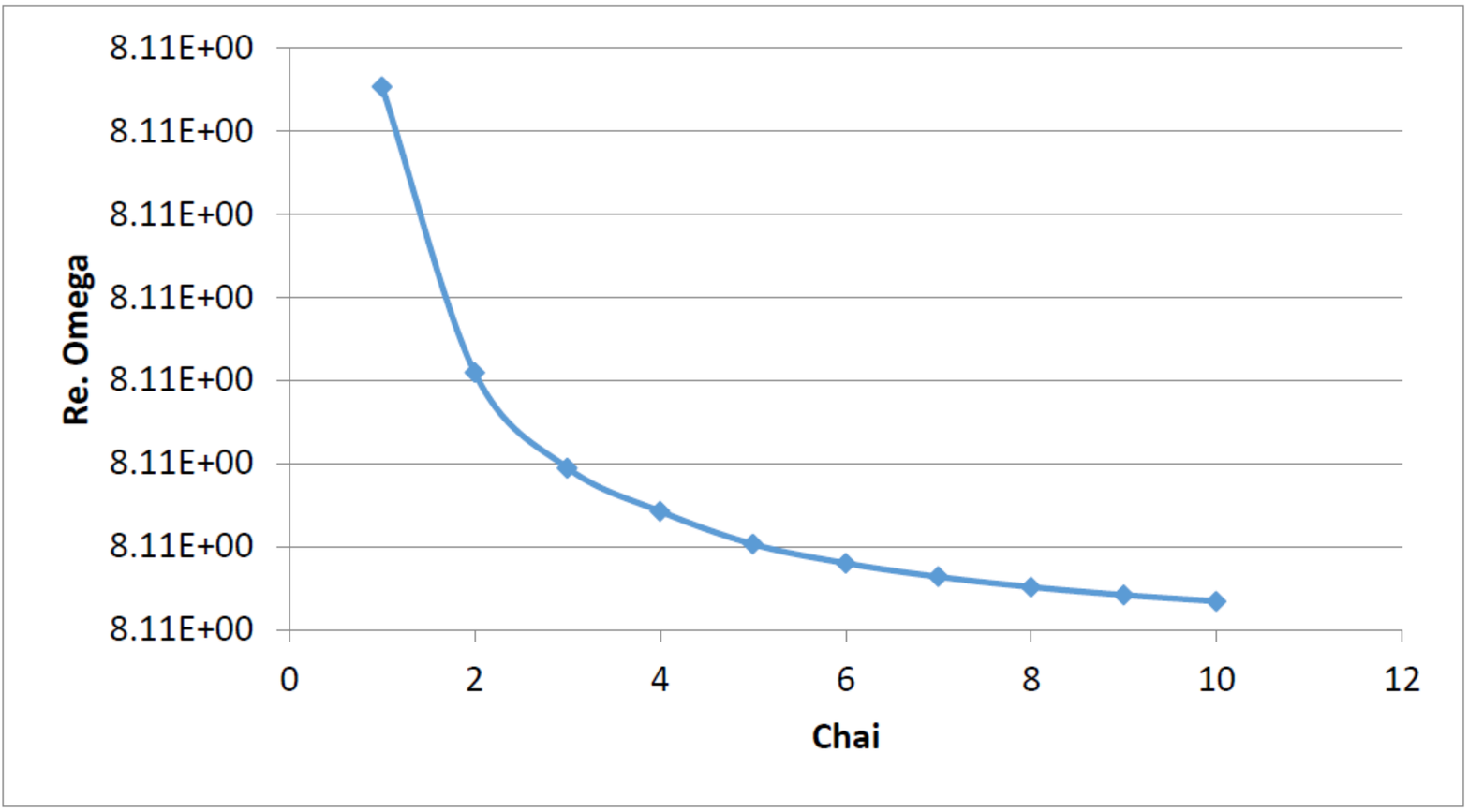}
			\caption{\label{fig:i}The imaginary part (left) and real part (right) of the quasinormal frequencies with respect to the thickness of the bulk, $\chi$ (Chai), for $Q = 2$ and $\zeta = 5 $. Here, we adopt $M=100$, $n = \kappa = 1$.}
		\end{center}
	\end{figure}
	\section{Quasinormal modes}
	Quasinormal modes are fundamental vibration modes around a black hole and they are obtained by solving equation (2.15) under the appropriate boundary condition. In order to consider boundary condition for an incoming wave at th event horizon, one can find that the effective potential $V\left( r \right) \to 0$ for $r \to {r_ + }$. However, for an outgoing wave at infinity, the effective potential vanishes, $V\left( r \right) = 0$, as the black hole tend to the Minkowski spacetime at infinity. Therefore, one can write the boundary condition for incoming and outgoing waves as
	\begin{equation}
	\left\{ \begin{array}{l}
	\psi \left( r \right) \sim {e^{ - i\omega {r^*}}}\begin{array}{*{20}{c}}
	{\begin{array}{*{20}{c}}
		{}&{}&{}
		\end{array}}&{as\begin{array}{*{20}{c}}
		{}&{}&{r \to {r_ + }\begin{array}{*{20}{c}}
			{}&{}&{\left( {{r^*} \to  - \infty } \right)}
			\end{array}}
		\end{array}}
	\end{array}\\
	\psi \left( r \right) \sim {e^{ + i\omega {r^*}}}\begin{array}{*{20}{c}}
	{\begin{array}{*{20}{c}}
		{}&{}&{}
		\end{array}}&{as\begin{array}{*{20}{c}}
		{}&{}&{r \to \infty \begin{array}{*{20}{c}}
			{}&{}&{\left( {{r^*} \to  + \infty } \right)}
			\end{array}}
		\end{array}}
	\end{array}
	\end{array} \right.
	\end{equation}
	In this manner, the general formalism to find the quasinormal modes of the black hole whose effective potential has the form of a potential barrier like that of formula (2.15), one can use WKB approximation which has been proposed by Schutz and Will \cite{schutz} , extended by Iyer and Will \cite{iyer} to the third order beyond the eikonal approximation, and subsequently developed by many people \cite{zhidenko,zhidenko1}. The WKB corrections up to sixth order have been used in \cite{lemos,zhidenko2,konoplya2}. In this case, when ${\left[ {V\left( r \right) - {\omega ^2}} \right]_{\max }} \ll \left[ {{\omega ^2} - V\left( { \pm \infty } \right)} \right]$, WKB has very good accuracy, i.e. if solutions of equation ${\omega ^2} - V\left( r \right) = 0$ are very close to each other as two turning points and the total energy, ${\omega ^2} - V\left( r \right)$ is expanded to the Taylor series near the maximum of the effective potential near by the turning points. In order to calculate the quasinormal modes \cite{iyer}, using the method called the $N$th order WKB approximation, one can get the formula for quasinormal frequencies as \cite{yoshida}
	\begin{equation}
	\omega  \simeq \sqrt {{V_0} + \sqrt {\frac{{V_0^{\left( 2 \right)}}}{2}} \left( {b + \frac{1}{2} + \sum\limits_{p = 0}^{N - 1} {{\Omega _p}} } \right)}
	\end{equation}
	where $V_0^{\left( p \right)}$ is the $p$th order derivative of the potential 
	\begin{equation}
	V_0^{\left( p \right)} \equiv {\left. {\frac{{{d^p}V\left( r \right)}}{{d{r^p}}}} \right|_{r = {r_0}}}
	\end{equation}
	and the first and the second of ${{\Omega _p}}$ are given by
	\begin{equation}
	\begin{array}{l}
	{\Omega _1} =  - 30{\left( {b + \frac{1}{2}} \right)^2}\gamma _1^2 + 6{\gamma _2}{\left( {b + \frac{1}{2}} \right)^2} - \frac{7}{2}\gamma _1^2 + \frac{3}{2}{\gamma _2},\\
	{\Omega _2} = {\left( {b + \frac{1}{2}} \right)^3}\left( { - 2820\gamma _1^4 + 1800\gamma _1^2{\gamma _2} - 280{\gamma _1}{\gamma _3} - 68\gamma _2^2 + 20{\gamma _4}} \right)\\
	\begin{array}{*{20}{c}}
	{}& + 
	\end{array}\left( {b + \frac{1}{2}} \right)\left( { - 1155\gamma _1^4 + 918\gamma _1^2{\gamma _2} - 190{\gamma _1}{\gamma _3} - 67\gamma _2^2 + 25{\gamma _4}} \right)
	\end{array}
	\end{equation} 
	Here, $\gamma_p$ for $(p \ge 1)$ is defined as follows
	\begin{equation}
	{\gamma _p} \equiv \frac{{V_0^{\left( {p + 2} \right)}}}{{\left( {p + 2} \right)!}}{\left( {\frac{1}{{2V_0^{\left( 2 \right)}}}} \right)^{\frac{p}{4} + 1}}.
	\end{equation}
	In this case, the parameter $b$ is called overtone number of quasinormal modes. We focused on $b = 0$ to obtain fundamental quasinormal modes as it is known that the approximation is good in case of $b < L$. We calculated the quasinormal frequencies for various values of the electromagnetic and tidal charges, $Q$ and $\zeta$, using the third order WKB method. Note that we fixed $n = \kappa = 1$ related to the quantum parameter $\lambda$ and considered a massless scalar field ($m = 0$). In the absence of the electromagnetic charge, $Q = 0$, in case of the $\zeta > 0$ the logarithmic imaginary part and real part of the $\omega$ decrease when $\zeta$ increases as well as when $\zeta < 0$ (figure 3). Here, we draw attention, in case of the absence of the electromagnetic and tidal charges, $Q = \zeta = 0$, it is not possible to calculate the quasinormal mode as ${\left. V \right|_{_{r = V_0^{\left( 1 \right)}}}} = 0$. Therefore, we excluded this point. We calculated the quasinormal frequencies for various values of $Q$ in the absence of the tidal charge, $\zeta = 0$. In this case, the imaginary part and real part of the $\omega$ decrease when the electromagnetic charge, $Q$ decreases (Figure 4). In Figure 5, we plotted quasinormal frequencies, both the real part and imaginary part, of the massless scalar field for the mass $M=100$ for various values of the electromagnetic charge. We took the angular parameter $L = 2$. In this case. we plotted a curve by changing $\zeta$ from -15 to 15. In this manner, both of the imaginary part and real part of the $\omega$ of the different values of the electromagnetic charge seems to converge to a value as we decrease the tidal charge $\zeta$. Obviously, the real part of the $\omega$ with respect to the imaginary part, has less variation when one increases the value of the electromagnetic charge, $Q$ (see Figure 6). We also calculated the quasinormal frequencies for different values of the thickness of the bulk in fixed value of the electromagnetic and tidal charges. It is also interesting to observe that the imaginary part of the quasinormal frequency increases when the thickness of the bulk increase while the real part of the $\omega$ decreases (Figure 7).

	\section{Conclusion and Discussion}
	In this paper, we have investigated quasinormal modes of massless scalar field of the charged black holes embedded in the Randall-Sundrum brane world. We found that the five dimensional effective potential $V$ depends on the value of $r$, integration constant related to $ADM$ mass, thickness of the bulk, $\chi$, and the electromagnetic and tidal charges, $Q$ and $\zeta$ respectively. In this case, we have computed the quasinormal frequencies spectrum of the charged black holes localized in the Randall-Sundrum brane world, using the third order of WKB approximation. 
	It is shown the quasinormal spectrum depends on the electromagnetic and tidal charges.
	We found, in the absence of the electromagnetic charge, $Q = 0$, the logarithmic imaginary part and real part of the $\omega$ decrease when $\zeta$ increases (figure 3). We also found the imaginary part and real part of the $\omega$ decrease when the electromagnetic charge, $Q$ decreases (Figure 4).
	We also have computed the quasinormal modes for different values of the electromagnetic and tidal charges in fundamental state $b  = 0$. We compare the real part and imaginary part of the quasinormal frequencies and their variations respect to the electromagnetic and tidal charges (Figure 5 and 6). We observed, that the imaginary part of the quasinormal frequency increases when the thickness of the bulk increase while the real part of the $\omega$ decreases (Figure 7).

	\section*{Acknowledgement}
	The paper is supported by Universiti Kebangsaan Malaysia (Grant No.FRGS/2/2013/ST02/UKM/02/1 for N. Abbasvandi). Shahidan Radiman would like to acknowledge receiving a grant DPP-2015-036 from UKM.

	%% The Appendices part is started with the command \appendix;
	%% appendix sections are then done as normal sections
	%% \appendix
	
	%% \section{}
	%% \label{}
	
	%% If you have bibdatabase file and want bibtex to generate the
	%% bibitems, please use
	%%
	%%  \bibliographystyle{elsarticle-num} 
	%%  \bibliography{<your bibdatabase>}

\begin{thebibliography}{00}
		
		%% \bibitem{label}
		%% Text of bibliographic item
		\bibitem{vishveshwara}
		C. V. Vishveshwara, \emph{Nature} {\bf 227} (1970) 936.
		
		\bibitem{chandrasekhar}
		S. Chandrasekhar, \emph{Proc. R. Soc. Lond. Ser. A.} {\bf 343} (1975) 289.	
		
		\bibitem{kokkotas}
		K. D. Kokkotas, B. G. Schmidt, \emph{Living Rev. Rel.} {\bf 2} (1999) 2.
		
		\bibitem{nollert}
		H. P. Nollert, \emph{Class. Quan. Grav.} {\bf 16} (1999) R159.
		
		\bibitem{konoplya}
		R. A. Konoplya, \emph{Phys. Rev. D} {\bf 68} (2003) 024018.
		
		\bibitem{berti}
		E. Berti, V. Cardoso, A. O. Starinets, \emph{Class. Quan. Grav.} {\bf 26} (2009) 163001.
		
		\bibitem{konoplya1}
		R. A. Konoplya, A. Zhidenko, \emph{Rev. Mod. Phys.} {\bf 83} (2011) 793.
		
		\bibitem{berti2003}
		E. Berti, K. D. Kokkotas, \emph{Phys. Rev. D} {\bf 68} (2003) 044027.
		
		\bibitem{berti20031}
		E. Berti, V. Cardoso, K. D. Kokkotas, and H. onozawa \emph{Phys. Rev. D} {\bf 68} (2003) 124018.
		
		\bibitem{berti2004}
		E.Berti, V. Cardoso, and S. Yoshida, \emph{Phys. Rev. D} {\bf 69} (2004) 124018.
		
		\bibitem{mann97}
		J. S. F. Chan, R. B. Mann, \emph{Phys. Rev. D} {\bf 55} (1997) 7546.
		
		\bibitem{mann98}
		J. S. F. Chan, R. B. Mann, \emph{Phys. Rev. D} {\bf 59} (1999) 064025.
		
		\bibitem{kalyanarama}
		S. Kalyana Rama, B. Sathiapalan, \emph{Mod. Phys. Lett.} {\bf A14} (1999) 2635.
		
		\bibitem{danielsson}
		U. H. Danielsson, E. Keski-Vakkuri, M. Kruczenski, \emph{JHEP} {\bf 02} (2000) 039.
		
		\bibitem{danielsson99}
		U. H. Danielsson, E. Keski-Vakkuri, M. Kruczenski, \emph{Nucl. Phys. B} {\bf 563} (1999) 279.
		
		\bibitem{brady}
		P. R. Brady, C. M. Chambers, W. G. Laarakkers, E. Poisson, \emph{Phys. Rev. D} {\bf 60} (1999) 064003.
		
		\bibitem{abdalla}
		E. Abdalla, C. Molina, A. Saa, \emph{gr-qc/0309078}.
		
		\bibitem{giammatteo}
		M. Giammatteo, I. G. Moss, \emph{Class. Quan. Grav.} {\bf 22} (2005) 1803.
		
		\bibitem{wang}
		B. Wang, C. Y. Lin, C. Molina, \emph{Phys. Rev. D} {\bf 70} (2004) 064025.
		
		\bibitem{chakrabarti}
		S. K. Chakrabarti, \emph{Gen. Rel. Grav.} {\bf 39} (2007) 567.
		
		\bibitem{lopez}
		A. Lopez-Ortega, \emph{Gen. Rel. Grav.} {\bf 38} (2006) 1747.
		
		\bibitem{cho}
		H. T. Cho, \emph{Phys. Rev. D} {\bf 68} (2003) 024003.
		
		\bibitem{jing04}
		J. Jing, \emph{Phys. Rev. D} {\bf 69} (2004) 084009.
		
		\bibitem{jing05}
		J. Jing, Q. Pan, \emph{Phys. Rev. D} {\bf 71} (2005) 124011.
		
		\bibitem{argyres}Argyres, P. C., Dimouplos, S., March-Russell, J., Phys. Lett. B, 441 96 (1998).  
		
		\bibitem{bank}Bank, T., Fischler, W., preprint hep-th/9906038
		
		\bibitem{emparan}Emparan, W., Horowitz, G. T., Myers, R. C., Phys. Rev. Lett., 85 499 (2000). 
		
		\bibitem{giddings}Giddings, S. B., Thomas, S., Phys. Rev. D, 77 045027 (2008). 
		
		\bibitem{arkani}Arkani-Hame, N., Dimoulos, S., Dvali, G., Phys. Lett. B, 429 263 (1998).  
		
		\bibitem{antoniadis}Antoniadis, I., Arkani-Hamed, N., Dimopoulos, S., Dvali, G., Phys. Lett. B, 436 257 (1998). 
		
		\bibitem{randall1}Randall, L., Sundrum, R., Phys. Rev. Lett., 83 3379 (1999). 
		
		\bibitem{randall2}Randall, L., Sundrum, R., Phys. Rev. Lett., 83 4690 (1999).
		
		\bibitem{kanti}
		P. Kanti, \emph{Int. J. Mod.  Phys. A} {\bf 19} (2004) 4899.
		
		\bibitem{kudoh}
		H. Kudoh, T. Tanaka, and T. Nakamura, \emph{Phys. Rev. D} {\bf 68} (2003) 024035.
		
		\bibitem{tanaka}
		T. Tanaka, \emph{Prog. Theor. Phys. Suppl.} {\bf 148} (2002) 307.
		
		\bibitem{fabbri}
		R. Emparan, A. Fabbri, and N. Kaloper, \emph{JHEP} {\bf 09} (2002) 043.	
		
		\bibitem{chung}
		H. Chung, L. Randall, M. J. Rodriguez, O. Varela, \emph{Classical Quantum Gravity} {\bf 33(24)} (2016) 245013.	
	
		\bibitem{ferrari}
		V. Ferrari, B. Mashhoon, \emph{Phys. Rev. D} {\bf 30} (1984) 295.
		
		\bibitem{leaver}
		E. W. Leaver, \emph{Proc. R. Soc. A} {\bf 402} (1985) 285.
		
		\bibitem{schutz}
		B. F. Schutz, C. M. Will \emph{Astrophy. J.} {\bf 291} (1985) L33.
		
		\bibitem{iyer}
		S. Iyer, C. M. Will, \emph{Phys. Rev. D} {\bf 35} (1987) 3621.
		
		\bibitem{chamblin}
		A. Chamblin, H.S. Reall, H. Shinkai, and T. Shiromizu, \emph{Phys. Rev. D} {\bf 63} (2001) 064015.
		
		
		\bibitem{kim}
		W. Kim, J.J. Oh, and Y.J. Park, \emph{Phys. Lett. B} {\bf 512} (2001) 131.
		
		\bibitem{zhidenko}
		R. A. Konoplya and A. Zhidenko, \emph{Nucl. Phys. B} {\bf 777} (2007) 182.
		
		\bibitem{zhidenko1}
		R. A. Konoplya, A. Zhidenko, \emph{Phys. Rev. D} {\bf 77} (2008) 104004.	
		
		
		\bibitem{shinkai}
		H. Shinkai, and T. Shiromizu, \emph{Phys. Rev. D} {\bf 62} (2000) 024010.
		
		\bibitem{liu}
		M. L. Liu, H. Y. Liu, L. X. Xu, and P. S. Wesson, \emph{Mod. Phys. Lett. A} {\bf 21} (2006) 2937.
		
		\bibitem{lemos}
		E. Berti, V. Cardoso, and J. P. S. Lemos \emph{Phys. Rev. D} {\bf 70} (2004) 124006.
		
		\bibitem{zhidenko2}
		A. Zhidenko, \emph{Class. Quan. Grav.} {\bf 21} (2004) 273.
		
		\bibitem{konoplya2}
		R. A. Konoplya, \emph{Phys. Rev. D} {\bf 68} (2003) 124017.
		
		\bibitem{yoshida}
		D. Yoshida, J. Soda, \emph{Phys. Rev. D} {\bf 93} (2016) 062003.
		
		
		
	\end{thebibliography}
	
	%% else use the following coding to input the bibitems directly in the
	%% TeX file.
	%\section*{References}

\end{document}